\documentstyle[aps,amsfonts,amssymb,prl,t1enc]{revtex}

\newcommand{\cH}{{\mathcal H}}

\newcommand{\cL}{{\mathcal L}}
\newcommand{\cM}{{\mathcal M}}

\newcommand{\cS}{{\mathcal S}}

\newcommand{\mb}{{\bar m}}

\newcommand{\sigmab}{{\bar\sigma}}
\newcommand{\taub}{{\bar\tau}}

\newcommand{\del}{\partial}

\newcommand{\thorn}{\mbox{\symbol{'376}}}
\renewcommand{\eth} {\mbox{\symbol{'360}}}
\newcommand{\2}{\frac12}

\title{On the Penrose inequality}
\author{J\"org Frauendiener}
\address{
Institut f\"ur Theoretische Astrophysik,\\
Universit\"at T\"ubingen,\\
Auf der Morgenstelle 10,\\
D-72076 T\"ubingen,\\
Germany}
\date{May 21, 2001}

\begin{document}
\maketitle

\begin{abstract}
The purpose of this letter is to point out an argument which may
ultimately lead to a rigorous proof of the Penrose inequality in the
general case. The argument is a variation of Geroch's original
proposal for a proof of the positive energy theorem which was later
adapted by Jang and Wald to apply to initial data sets containing
apparent horizons. The new input is to dispense with the a priori
restriction to an initial data set and to use the four-dimensional
structure of spacetime in an essential way.
\end{abstract}
\pacs{04.20.-q,02.40.-k}

In an attempt to find a counterexample for the cosmic censorship
hypothesis Penrose~\cite{penrose73:_naked_sing} postulated an
inequality which relates the area of an apparent horizon to the total
mass of an isolated gravitating system. His reasoning, based on what
he called the ``establishment view'' on gravitational collapse, was
roughly as follows: consider an asymptotically flat spacetime
containing mass/energy at sufficiently high densities that it
collapses under its own gravitational attraction so that during the
collapse a marginally trapped surface $\cH$ and later a trapped
surface forms. The singularity theorem implies that the spacetime will
develop a singularity in the future. If the cosmic censorship
hypothesis is valid, then there will exist an event horizon which
encloses the ensuing singularity and the surface~$\cH$. Following the
``establishment view'' on gravitational collapse Penrose argues that
the spacetime will ultimately settle down to a Kerr black-hole. Then
the area $A_{\cH}$ of the apparent horizon $\cH$ will be less than the
area of the intersection of the event horizon with any spacelike and
asymptotically flat hypersurface containing $\cH$. Since the area of
the event horizon increases towards the future $A_{\cH}$ will also be
less than the area of the horizon in the limiting Kerr solution which
is given by the well known formula $A_{Kerr} = 8\pi m\left(m
+\sqrt{m^2 - a^2} \right)$ which in turn is less than or equal to the
Schwarzschild value $A_{S}=16\pi m^2$ for the same mass. This mass,
however, cannot exceed the value of the total (ADM) mass $M$ of the
system because the Bondi-Sachs mass loss formula implies that during
the collapse gravitational radiation had carried away positive energy
towards null-infinity. Thus, the final inequality is
\begin{equation}
A_{\cH} \le 16\pi M^2.\label{eq:penineq}
\end{equation}
If one could set up situations in which this inequality was violated
then this would provide a strong argument against the validity of the
cosmic censorship hypothesis. Although, on the other hand, an
independent proof of this inequality would not be a proof of cosmic
censorship, there have been numerous attempts to show that the Penrose
inequality~(\ref{eq:penineq}) is, in fact, true without referring to
the cosmic censorship hypothesis. One of the most successful attempts
has been the argument by Jang and
Wald~\cite{jangwald77:_pos_energy_conj_cch} which is based in turn on
Geroch's earlier idea~\cite{geroch73:_energ} to prove the positive
energy theorem.

Note, that the inequality~(\ref{eq:penineq}) relates the area of a
finite spacelike 2-surface in spacetime with a quantity defined at
infinity. Thus, it is natural to assume that $\cH$ is contained in an
initial data surface $\Sigma$ which extends out to (spacelike)
infinity. Then the problem of proving~(\ref{eq:penineq}) becomes a
problem within the context of initial data sets for the Einstein
equations. The basic idea in~\cite{jangwald77:_pos_energy_conj_cch}
was to use the Hawking mass of the 2-surfaces of a foliation of
$\Sigma$ as a quantity which interpolates between the area of the
apparent horizon on the one hand and the total mass on the other
hand. Geroch~\cite{geroch73:_energ} had shown that, with some
technical assumptions, the rate of change of the Hawking mass along
successive 2-surfaces is positive definite provided that the dominant
energy condition holds and that the 2-surfaces evolve according to the
inverse mean curvature flow. That means that in order to transform one
surface into the next each of its points is moved a distance
proportional to the inverse of the mean curvature at that point in the
direction of the normal at that point.  The most limiting assumption
in this argument is that the initial data be time symmetric, i.e. its
extrinsic curvature in spacetime should vanish. Thus, if the inverse mean
curvature flow existed and was smooth then the reasoning
in~\cite{jangwald77:_pos_energy_conj_cch} would provide a proof of the
Penrose inequality~(\ref{eq:penineq}) for time symmetric initial data
sets. This case, the so called ``Riemannian Penrose inequality'', was
recently proven rigorously by Huisken and
Ilmanen~\cite{huiskenilmanen97:_rieman_penros,huiskenilmanen98:_rieman_penros}
by showing the existence of an appropriate weak version of the inverse
mean curvature flow which suffices to make the argument work.
However, there still remains the fact that $\Sigma$ has to satisfy
additional properties so that the general proof of the Penrose
inequality is still lacking.

As noted above, the inequality relates the finite 2-surface $\cH$ and
infinity. These are the ``given data'' while the initial hypersurface
$\Sigma$ is added by hand. A priori, no spacelike hypersurface
$\Sigma$ is preferred. This raises the question as to whether it is
possible to use the available data to \emph{construct} an appropriate
$\Sigma$. An indication of how this might be achieved can be found in
the work of
Israel~\cite{israel86:_black_holes_cch,israel86:_confinement_theorem}
(see also Needham~\cite{needham86:_two_probl_cch}). He considered a
trapped 2-surface and showed how to extend it into the future by a
3-dimensional spacelike cylinder whose sections are all trapped so
that its interior is causally incarcerated. 

Consider a spacelike 2-surface $\cS$ embedded in spacetime
$\cM$\footnote{We will use the notation and conventions of Penrose and
Rindler~\cite{penrose86:_spinor_spacet_ii} throughout.}. Associated
with $\cS$ are two unique null-directions perpendicular to $\cS$. Let
$l^a$ and $n^a$ be two null-vectors aligned along these respective
null-directions normalized against each other and let $\rho$ and
$\rho'$ be the divergences of the respective null-geodesic congruences
emanating from $\cS$. Clearly, the null-vectors are only defined up to
scale. Under a rescaling $l^a \mapsto c l^a$ the other null-vector and
the divergences scale according to $n^a \mapsto (1/c) n^a$, $\rho
\mapsto c \rho$ and $\rho' \mapsto (1/c) \rho'$. Thus the combinations
$\rho' l^a$ and $\rho n^a$ are invariantly defined as are all their
linear combinations. Of special interest are the combinations $V^a =
\rho'l^a - \rho n^a$ (which was used by Israel and Needham) and $H^a =
\rho' l^a + \rho n^a$. In ``normal situations'' such as for a convex
surface in Minkowski space when one of the null congruences diverges
while the other converges i.e., when $\rho < 0$ and $\rho' > 0$ then
$V^a$ and $H^a$ are timelike and spacelike, respectively. The vector
field $H^a$ is, in fact, the \emph{mean curvature vector} of $\cS$ in
$\cM$. We define the \emph{inverse mean curvature vector} by (with
$H^2 = -H^aH_a$)
\begin{equation}
  \label{eq:mcv}
  q^a = \frac1{H^2}H^a = -\frac1{2\rho} l^a - \frac1{2\rho'} n^a.
\end{equation}
We can use $q^a$ on $\cS$ to construct another (infinitesimally close)
2-surface $\cS'$ according to the rule that we move every point of
$\cS$ a unit distance along the vector $q^a$ at that point. From
$\cS'$ we proceed according to the same rule and so on. Thus, we
ultimately construct a family of spacelike 2-surfaces $\cS_\lambda$
which are linked along a spacelike vector field $q^a$ with parameter
$\lambda$ to form a spacelike 3-dimensional hypersurface $\Sigma$.

Next, consider the Hawking mass~\cite{hawking68:_gravit_radiat}
associated with the 2-surfaces $\cS_\lambda$
\begin{equation}
  \label{eq:hawkingmass}
  m[\cS_\lambda] = \frac1{4\pi}\left(\frac{A}{4\pi}\right)^{1/2} \int_{\cS_\lambda}
  \left[ K + \rho\rho' \right]\, d^2S.
\end{equation}
Here $K$ is the so called complex curvature of the 2-surface
$\cS_\lambda$. As a consequence of the Gau\ss-Bonnet theorem its
integral over the surface evaluates to a real multiple of the Euler
characteristic of the surface.  We are interested in the rate of
change of $m[\cS_\lambda]$ along the surfaces. Since we are
considering spacelike 2-surfaces we apply the
GHP-formalism~\cite{gerochheld73:_ghp,penrose84:_spinor_spacet_i}. On
each member $\cS_\lambda$ the two null-directions are fixed and we
choose null-vectors $l^a$ and $n^a$ along them. The complex spacelike
null-vector $m^a$ is tangent to $\cS_\lambda$ and defined up to $m^a
\mapsto \gamma m^a$ with $\gamma\bar\gamma=1$. Since $m^a$ is
Lie-transported along $q^a$ up to this indeterminacy we have the
equation $\cL_q m^a = A m^a + B \mb^a$ for some complex-valued
functions $A$ and $B$ on $\cS_\lambda$. This equation can be rewritten
in the form
\begin{equation}
  \label{eq:nablaqm}
  q^a\nabla_a m^c = \eth q^c + A m^c + B \mb^c
\end{equation}
from which  we can derive the relationship
\begin{equation}
  \label{eq:ethrho}
  \eth\rho + \tau \rho + \rho' \kappa =0
\end{equation}
between the spin-coefficients $\rho$, $\rho'$, $\kappa$ and $\tau$
together with its complex conjugate and primed versions. The change of
the area along the inverse mean curvature vector is given by
\begin{equation}
  \label{eq:darea}
  \frac{d}{d\lambda}A[\cS_\lambda] = \frac{d}{d\lambda}
  \int_{\cS_\lambda} d^2A = \int_{\cS_\lambda} \cL_q 
  d^2A = 2A[\cS_\lambda],
\end{equation}
from the definition of the divergences $\rho$ and $\rho'$. 

The change in the Hawking mass is 
\begin{eqnarray}
  \label{eq:dmass}
  \frac{d}{d\lambda}&& m_H[\cS_\lambda] 
  =  m_H[\cS_\lambda] \nonumber\\
&&+ \frac1{4\pi}\left(\frac{A}{4\pi}\right)^{1/2}
  \int_{\cS_\lambda} \left[2 \rho \rho'+ \cL_q  \left( \rho \rho'\right) \right]  d^2A. 
\end{eqnarray}
From the definition of $q^a$ and the appropriate GHP equations we
have\footnote{Since the result of this computation cannot depend
on how the null-vectors are extended off the surfaces, we assume that
they are geodesic, i.e., $\kappa=\kappa'=0$ on $\cS_\lambda$. The
calculation without this assumption yields the same result.}
\begin{eqnarray*}
  \label{eq:Lqrho2}
 \cL_q&&\left( \rho \rho'\right)  = \rho' q^a \nabla_a \rho + \rho q^a
 \nabla_a \rho'
 = -\2 \frac{\rho'}{\rho} \thorn \rho - \2 \thorn' \rho + \mbox{``primed''}\\
 &&= -\2 \frac{\rho'}{\rho} \left\{ \rho^2 + \sigma\sigmab +
   \Phi_{00}\right\} - \\
 && - \2 \left\{ \eth'\tau + \rho\rho' + \sigma\sigma' - \tau \taub -
   \Psi_2 - 2\Lambda \right\} + \mbox{``primed''}
\end{eqnarray*}
Here, ``primed'' indicates the terms obtained from the displayed ones
by the priming operation
(see~\cite{penrose86:_spinor_spacet_ii}). Using the definition of the
complex curvature
\[
K = \sigma\sigma' - \Psi_2 - \rho\rho' + \Phi_{11} + \Lambda 
\]
and the Einstein equations in the form
\begin{equation}
  \label{eq:einstein}
  \Phi_{ab} = 4\pi G\,\left(T_{ab} - \frac14 g_{ab}\, T^c{}_c\right),
  \quad \Lambda = \frac13 \pi G\,T^c{}_c
\end{equation}
yields
\begin{eqnarray*}
  \cL_q&&\left( \rho \rho'\right) = -3 \rho\rho' - K + \2 \left(
  -\eth'\tau - \eth \tau' + \tau \taub + \tau' \taub' \right) + \\ &&
+ \frac1{H^2} \left\{ \rho'^2 \sigma\sigmab + \rho^2 \sigma'\sigmab' +
  4\pi G\, T_{ab} V^a V^b \right\}
\end{eqnarray*}
Inserting this into~(\ref{eq:dmass}) and using~(\ref{eq:ethrho}) and
its primed version to rewrite the $\tau$-terms in terms of the
divergences $\rho$ and $\rho'$ (under the simplifying but not
restricting assumption that $\kappa=\kappa'=0$) gives the final result
\begin{eqnarray}
  \frac{d}{d\lambda} m_H[\cS_\lambda] = \2 &&\int_{\cS_\lambda} \left\{
    \frac{\eth\rho}{\rho}\frac{\eth'\rho}{\rho} +
    \frac{\eth\rho'}{\rho'}\frac{\eth'\rho'}{\rho'}  \right\}\,d^2A \nonumber \\
  + &&\int_{\cS_\lambda} \frac1{H^2} \left\{ \rho'^2 \sigma\sigmab +
    \rho^2 \sigma'\sigmab' \right\} \,d^2A  \\
  \label{eq:dmassfinal}
  +   4\pi G\, &&\int_{\cS_\lambda} \frac1{H^2} T_{ab} V^a V^b \, d^2A
\nonumber
\end{eqnarray}
Thus, as long as $\rho$ and $\rho'$ remain non-zero, the rate of
change of the Hawking mass along the foliation of 2-surfaces consists
of three terms which are manifestly positive provided the dominant
energy condition is valid. Note, that the second and third terms are
associated with the energy flux penetrating through each particular
2-surface. The third term is the density of the matter as measured by
an observer moving along the timelike vector $V^a$ perpendicular to
$q^a$ (note that $V^aV_a=H^2$) while the second term can be
interpreted as the flux of gravitational wave energy along the two
distinguished null-directions~\cite{penrose66:_gener_relat}. The
physical meaning of the first term is unclear. There seems to be a
superficial similarity to the expression for the Newtonian
gravitational energy density $\del_a\phi \del^a\phi$ if the
gravitational potential $\phi$ is formally replaced with $\log\rho$ or
$\log\rho'$. This analogy is somewhat supported by explicit examples
in Schwarzschild spacetime but this needs to be explored further.

As the 2-surfaces flow out along the inverse mean curvature vector
they generate a spacelike hypersurface and the question we need to ask
is where this hypersurface will ultimately end up. Will it necessarily
become asymptotically Euclidean or could it happen that the
hypersurface becomes hyperboloidal i.e. that it bends ``up'' or
``down'' just enough to approach null-infinity while remaining
spacelike throughout? Since there is no preference in the setup for
one or the other time direction this is difficult to imagine and
preliminary studies seem to confirm the belief that the hypersurface
will become asymptotically Euclidean. However, this is a very difficult
question which can probably be answered only by the full existence
proof for the flow.

Suppose that spacetime contains a marginally trapped surface $\cH$
with spherical topology and assume that there is a spacelike
hypersurface $\Sigma$ foliated by 2-surfaces
$\{\cS_\lambda\}_{\lambda\ge0}$ obtained by flowing along the inverse
mean curvature vector $q^a$. If $\Sigma$ is such that $\cS_0=\cH$ and is
asymptotically Euclidean then
\[
\lim_{\lambda\to\infty} m[\cS_\lambda] = M
\]
i.e. the limit of the Hawking mass at spacelike infinity is the ADM
mass. Furthermore, at the apparent horizon we have $\rho=0$ and
\[
m[\cH] = \frac12\sqrt{\frac{A}{4\pi}}.
\]
Since the Hawking mass never decreases along the inverse mean curvature
vector we have $m[\cH] \le M$ or, equivalently, $A_{\cH} \le 16\pi M^2$.

The present argument is very similar to the Jang-Wald argument in that
the Hawking mass is used as the principal tool to relate the two
quantities in question while an appropriate flow provides the
foliation along which the mass increases. The main difference,
however, is that here the flow itself defines the spacelike
hypersurface. The hypersurface ``grows'' towards infinity in a manner
which is dictated by the energy contents of the spacetime so that the
mass never decreases. When viewed within the generated hypersurface
$\Sigma$, the vector field $q^a$ coincides with (twice) the inverse
mean curvature vector of the 2-surfaces $\cS_\lambda$ in $\Sigma$. The
extrinsic curvature of $\Sigma$ has the peculiar property that its
restriction to any of the leaves $\cS_\lambda$ is tracefree, in other
words, the mean curvature of $\Sigma$ is equal to the negative normal
component $K_{ab} r^ar^b$ with $r^a$ the normal to $\cS_\lambda$. This
is the necessary and sufficient condition for writing $\rho\rho' =
-p^2/8$ with $p$ the mean curvature of $\cS_\lambda$ in $\Sigma$. With
this and the relationship $\mbox{Re}(K) = {}^2R/4$ between the scalar
and complex curvatures of the 2-surfaces we obtain
\[
\frac18\left[ 2 {}^2R - p^2 \right]
\]
for the integrand in the mass integral~(\ref{eq:hawkingmass}) which
completely agrees with the expression used by Jang and Wald.

Despite the similarity of the arguments the technical problems seem to
be completely different. While the Jang-Wald case was concerned with
flowing 2-surfaces in a three-dimensional Riemannian manifold, the
present situation has to handle the flow of 2-surfaces in a
four-dimensional (Lorentzian) spacetime. Obviously, the present
argument does not (yet) provide a rigorous proof of the Penrose
inequality but it gives a mathematically beautiful and physically
convincing demonstration of the fact that the origin of the inequality
lies in the four-dimensional spacetime structure.

The main question which remains to be answered is that of existence
and regularity of the flow. This is the same situation as with the
Jang-Wald argument which has been made rigorous by Huisken and
Ilmanen. Just like in that case there is no reason to assume that the
flow will remain smooth. If at any point one of the divergences
vanishes then the surface will shoot off with infinite speed towards
null-infinity along one of the null-directions. This will result in a
discontinuous jump in the flow and the question is whether one can
find a formulation for the flow problem which can cope with such
situations while still allowing for the above argument. It should be
noted also that the starting point of the evolution for the flow is
singular because the vector field is singular on the apparent horizon
just as in the Riemannian case where the inverse mean curvature vector
is singular at the initial minimal surface.  However, the hope is that
just as in the Riemannian case one can reformulate the flow problem in
a way which allows for an appropriate regularization of the flow. Then
it might also be possible to start with a trapped surface instead of a
marginally trapped one and also to deal with the case when the
starting surface has several components.

This work grew out of the attempt to understand the peculiar
properties of certain integral
formulae~\cite{frauendiener97:_integr_formula} which provide a proof
of the Penrose inequality in various special cases. This relationship
will be discussed in more detail in a future paper.

I wish to thank R. Penrose for drawing my attention to
ref.~\cite{needham86:_two_probl_cch} and to T. Needham for making it
available to me. Also I am grateful to K. P. Tod and L. B. Szabados
for valuable comments on an earlier draft of this paper.


\begin{thebibliography}{10}

\bibitem{penrose73:_naked_sing}
R. Penrose, Ann. N.Y. Acad. Sci. {\bf 224},  125  (1973).

\bibitem{jangwald77:_pos_energy_conj_cch}
P.~S. Jang and R. Wald, J. Math. Phys. {\bf 18},  41  (1977).

\bibitem{geroch73:_energ}
R.~P. Geroch, Ann. N. Y. Acad. Sci. {\bf 224},  108  (1973).

\bibitem{huiskenilmanen97:_rieman_penros}
G. Huisken and T. Ilmanen, Int. Math. Res. Not. {\bf 20},  1045  (1997).

\bibitem{huiskenilmanen98:_rieman_penros}
G. Huisken and T. Ilmanen, to appear in J. Diff. Geom.

\bibitem{israel86:_black_holes_cch}
W. Israel, Can. J. Phys. {\bf 64},  120  (1986).

\bibitem{israel86:_confinement_theorem}
W. Israel, Phys.~Rev.~Lett. {\bf 56},  789  (1986).

\bibitem{needham86:_two_probl_cch}
T. Needham, Ph.D. thesis, Wolfson College, Oxford, 1986.

\bibitem{penrose86:_spinor_spacet_ii}
R. Penrose and W. Rindler, {\em Spinors and Spacetime} (Cambridge University
  Press, Cambridge, 1986), Vol.~2.

\bibitem{hawking68:_gravit_radiat}
S.~W. Hawking, J. Math. Phys. {\bf 9},  598  (1968).

\bibitem{gerochheld73:_ghp}
R.~P. Geroch, A. Held, and R. Penrose, J. Math. Phys. {\bf 14},  874  (1973).

\bibitem{penrose84:_spinor_spacet_i}
R. Penrose and W. Rindler, {\em Spinors and Spacetime} (Cambridge University
  Press, Cambridge, 1984), Vol.~1.

\bibitem{penrose66:_gener_relat}
R. Penrose,  in {\em Perspectives in Geometry and Relativity}, edited by B.
  Hoffmann (Indiana University Press, 1966), pp.\ 259--274.

\bibitem{frauendiener97:_integr_formula}
J. Frauendiener, Class. Quant. Grav. {\bf 14},  3413  (1997).

\end{thebibliography}

\end{document}